# PULSED LASER DEPOSITION OF $YBa_2Cu_3O_{7-\delta}$ FOR COATED CONDUCTOR APPLICATIONS: CURRENT STATUS AND COST ISSUES

Hans M. Christen

Oak Ridge National Laboratory
Solid-State Division
Oak Ridge, TN 37831-6056

## INTRODUCTION

Amongst the numerous techniques currently being tested for the fabrication of coated conductors (i.e. high-temperature superconducting oxides deposited onto metallic tapes), pulsed laser deposition (PLD) plays a prominent role. Recent results from groups in the United States (e.g. Los Alamos National Laboratory), Europe (e.g. University of Göttingen), and Japan (e.g. Fujikura Ltd.) are most promising, yielding record numbers for $J_c$ and $I_c$.

As a method for depositing films of complex materials, such as the high-temperature superconductors (HTS) relevant to this chapter, PLD is well established and conceptually simple. Issues that have kept PLD from becoming a successful technique for devices and optical materials, namely particulate formation and thickness non-uniformities, are much less of a concern in the fabrication of HTS tapes. Nevertheless, the approach still presents ongoing challenges in the fundamental understanding of the laser-target interaction and growth from the resulting energetic plasma plume. Numerous issues related to the scale-up, control, reproducibility, and economic feasibility of PLD remain under investigation.

It is clearly beyond the scope of this short chapter to give a complete treatise of such a complex subject – fortunately, several excellent reviews have already been published.[1-3] It is thus the intent of the author to introduce PLD only briefly and with a strong focus on HTS deposition, summarizing the most important developments and referring the reader to the numerous cited works. A strong emphasis is placed on issues related to scale-up.

Finally, it is the goal of this chapter to introduce a preliminary cost-analysis of PLD for coated conductor fabrication. While an attempt to calculate the total cost of fabricating a length of superconducting tape would be beyond the scope of this chapter (and beyond our current knowledge of numerous issues), we can present a rather detailed analysis of the cost of laser operation. The point, of course, is not to show that PLD is either cheaper or more expensive than other techniques – in fact, for most other techniques, which have not been successfully applied to long length fabrication, such a calculation cannot yet be made. It is illustrative, for example, to consider that in an industrial setting, more meters of coated tape will have to be produced *per day* than some other methods have produced in their entire history, and that therefore many reliability issues (including mean-time between failure of equipment) have not been addressed for many of the approaches described in this book. It is the maturity of industrial laser technology combined with the maturity of the PLD process that allows us to estimate these costs for this technique.

## BASIC PRINCIPLES OF PLD

### History and fundamental mechanisms

The use of a pulsed laser to induce the stoichiometric transfer of a material from a solid source to a substrate, simulating the flash evaporation methods that have previously been successful, is reported in the literature as early as 1965,[4] where films of semiconductors and dielectrics were grown using a ruby laser. Pulsed laser evaporation for film growth from powders of $SrTiO_3$ and $BaTiO_3$ was achieved in 1969.[5] Six years later, stoichiometric intermetallic materials (including $Ni_3Mn$ and low-$T_c$ superconducting films of $ReBe_{22}$) were produced using a pulsed laser beam.[6] In 1983, Zaitsev-Zotov and co-workers for the first time reported superconductivity in pulsed laser evaporated oxide superconductor films after heat-treatment.[7]

The real breakthrough for PLD, however, was its successful application to the in-situ growth of epitaxial high-temperature superconductor films in 1987 at Bell Communications Research.[8]

Since then, PLD has been used extensively in the growth of high-temperature cuprates and numerous other complex oxides, including materials that cannot be obtained in an equilibrium process.

Conceptually, the process of PLD is extremely simple and illustrated schematically in Fig. 1. A pulsed laser beam leads to a rapid removal of material from a solid target and to the formation of an energetic plasma plume, which then condenses onto a substrate. In reality, however, the individual steps – ablation and plasma formation, plume propagation, and nucleation and growth – are rather complex.

**Ablation and plasma formation.** Ablation has been studied extensively, not only in connection to PLD, but also because of its importance in laser machining. The mechanisms that convert the electromagnetic energy of the coherent light beam first into electronic excitations and then into thermal, chemical, and mechanical energy are complex[9,10] and still not fully understood. Heating rates as high as $10^{11}$ K$^{-1}$ and instantaneous gas pressures of 10 – 500 atm. are observed at the target surface.[11] The laser-solid interaction mechanisms depend strongly on the laser wavelength; in fact, significant changes in the energetics of species in a plume resulting from ablation of carbon using KrF (248nm) and ArF (193 nm) excimer lasers are observed,[12] having a large effect on the growth of diamond-like carbon films. Quite surprisingly, some ablated surfaces show topographic features oriented along the polarization axis of the incident laser beam rather than the beam direction after ablation using a femtosecond laser,[13] illustrating that for very short pulses, thermal effects are not sufficient to describe the process.

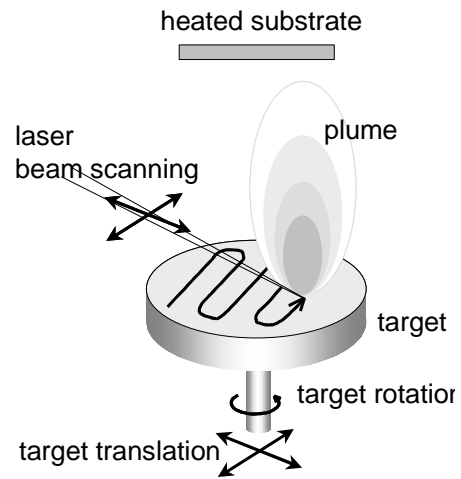

**Figure 1.** Schematic drawing of the basic components inside a vacuum system as used in pulsed laser deposition. The laser beam is scanned across a target, which may be a cylindrical disc, but could have other shapes, such as a cylindrical rod with ablation from the side.

For relatively long pulse durations, such as the tens of nanoseconds typical for excimer lasers, there is a strong interaction between the forming plume and the incident beam, leading to a further heating of the species. This may explain experiments of $YBa_2Cu_3O_{7-\delta}$ film growth where, for a given laser energy density at the target surface, ablation using a KrF excimer laser (248 nm, ≈ 30 ns pulse duration) resulted in far superior films than ablation using a frequency-quadrupled Nd:YAG (266 nm, ≈ 5 ns pulse duration).[14] Similarly, certain aspects of a dual-laser approach,[15] where a $CO_2$ laser pulse with a 500 ns duration is allowed to interact with the plume formed by the ablation using a KrF excimer laser, have been attributed to increased laser heating of the plasma.

**Plume propagation.** Extensive experiments have been performed to study the plume propagation using optical absorption and emission spectroscopy combined with ion probe measurements.[11,16] Neutral atoms, ions, and electrons travel at different velocities, and strong interactions between the species of the plasma and the background gas are observed. In fact, it is generally assumed that some degree of thermalization needs to occur in order to obtain good film growth and to avoid resputtering of the growing film by the most energetic ions in the plume.[17] Assuming that most of the species in the plume should be fully thermalized at precisely the time they reach the substrate (i.e. having equal lateral and forward velocities), a simple model predicts that the optimal growth rate should be close to 1 Å per pulse.[18,19] This is rather close to the actually observed values (typically ranging from 0.1 Å to 1Å per pulse).

In excimer laser ablation experiments of $YBa_2Cu_3O_{7-\delta}$, the formation of nanoparticles in the plume has been observed at oxygen pressures above 175 mTorr and at room temperature.[20] However, for a heated substrate, the temperature gradient moves nanoparticles away from the heater surface, such that in typical $YBa_2Cu_3O_{7-\delta}$ growth experiments, nanoparticle incorporation into the growing films appears not to be an issue.

**Nucleation and growth.** Detailed descriptions of the growth modes observed in PLD have been published.[21-24] Island growth occurs most frequently, but layer-by-layer growth can occur at sufficiently low growth rates and sufficiently elevated temperatures. Recent experiments indicate that in general, crystallization is fast (as observed using time-resolved surface x-ray diffraction[25]), but changes at the surface are observed for times ranging from factions of a second to several seconds depending on the growth conditions (as observed both in RHEED[26] and in surface x-ray diffraction[25]). At very high temperature and low pressure, step flow growth was observed on some oxide materials.[27] However, for the fabrication of long lengths of HTS tapes, these slower growth modes are neither achievable nor necessary.

The term <u>Laser-MBE</u> has been used to describe a PLD system in which layer-by-layer growth is achieved and monitored by RHEED. The terminology, of course, is somewhat inaccurate, as a laser plume always contains a combination of ions, electrons, and neutral particles and is thus not a molecular beam. Nevertheless, "laser-MBE" has been successfully used to sequentially deposit single layers of SrO and BaO (Ref. 28) and to intercalate SrO layers in manganites to form artificial crystalline structures.[29] The term has further been used when ablation occurred from a complex target,[30-32] thus "laser-MBE" is often synonymous with "layer-by-layer growth by PLD."

**Laser requirements.** The key parameters for a laser to be used in PLD are its wavelength, pulse duration, and energy per pulse. A sufficiently small wavelength assures that most of the energy is absorbed in a very shallow layer near the surface of the target; otherwise subsurface boiling can occur, leading to a large number of particulates at the film surface. The absorption of photons by oxygen molecules and optical elements in the beam path determines a lower practical wavelength limit of about 200 nm. The pulse duration must be short enough such as not to lead to a significant heating of the bulk of the target (again to avoid boiling, particulate formation, and changes in stoichiometry at the surface), but long enough to transfer some energy into the plasma. Finally, the laser energy at the surface of the target has to exceed a certain threshold, typically $1 - 3$ J/cm$^2$ for a 30 ns pulse. Therefore, the laser energy per pulse will determine the spot size over which the laser must be focused, and thus the amount of material that can be ablated per pulse. In addition, it is generally observed that a small laser spot results in a broad angular distribution of the ablated species, so that it is impossible to change the energy per pulse without significant other modifications in the growth apparatus.

Excimer lasers appear to satisfy all of these requirements (see Ref. 33 for an excellent introduction to excimer lasers). Lasers using KrF excimers (248 nm, typically $20 - 35$ ns pulse duration) have been used most often in PLD. Successful YBa$_2$Cu$_3$O$_{7-\delta}$ film growth has also been achieved using ArF (193 nm)[34-36] and XeCl (308 nm)[37-41] excimers.

For reasons mentioned above, Nd:YAG lasers have only rarely been used for the growth of YBa$_2$Cu$_3$O$_{7-\delta}$.[42]

The growth of less complex materials has been possible using a variety of lasers, including hybrid dye/excimer lasers (248 nm, 500 fs) for amorphous carbon nitride,[43] femtosecond Ti-sapphire lasers for ZnO,[44,45] SnO$_2$,[46] carbon[47,48] and AlN.[49] A method of "ultrafast ablation" has been proposed and applied to the growth of amorphous carbon.[50] Here, either a 10 kHz, 120 ns Q-switched Nd:YAG laser or a 76 MHz, 60 ps mode-locked Nd:YAG laser are used, resulting in very smooth films at higher growth rates than in conventional PLD, but again the approach has not successfully been applied to YBa$_2$Cu$_3$O$_{7-\delta}$.

**PULSED LASER DEPOSITION OF YBa$_2$Cu$_3$O$_{7-\delta}$**

Despite the fact that the first YBa$_2$Cu$_3$O$_{7-\delta}$ films were grown almost 15 years ago, new achievements and results are published continuously. For example, very recent publications indicate that it is now possible to grow double-sided YBa$_2$Cu$_3$O$_{7-\delta}$ reproducibly on large numbers of 3" diameter substrates (for filter applications),[51] that YBa$_2$Cu$_3$O$_{7-\delta}$ can be grown onto flexible yttria-stabilized zirconia (YSZ) tapes for cryoelectronic applications,[52] and that silver doping can result in higher J$_c$ values.[53]

Pulsed laser deposition has been used extensively in research efforts related to coated conductors. Numerous authors have reported critical current densities in excess of $10^6$ A/cm$^2$.[37,54-59] PLD-grown films have contributed to studies on the effect of texture in coated conductors,[59,60] ac loss experiments,[61] bend strain tolerance,[55] and, in many studies, as a means to quantify the quality of buffer layers grown by a variety of methods.[58,62-67]

**Recent results for long lengths** of coated conductors are promising.[*] Reel-to-reel deposition of $YBa_2Cu_3O_{7-\delta}$ with end-to-end $I_c > 10$ A for a 4.5 m tape and $I_c = 62$ A (1 cm tape width) for shorter lengths and a film thickness of 4 μm were reported by Sato *et al.*[68] Y. Iijima's group (Fujikura Ltd.) reported on a 9.6 m long carrying an $I_c = 50$ A (1 cm tape width), corresponding to a current density $J_c = 0.42$ MA/cm$^2$. On a shorter length of 8 cm (for which the buffer layer was deposited at a slower rate), $I_c = 140$ A (1 cm tape width, corresponding to $J_c = 1.2$ MA/cm$^2$) was measured.[69]

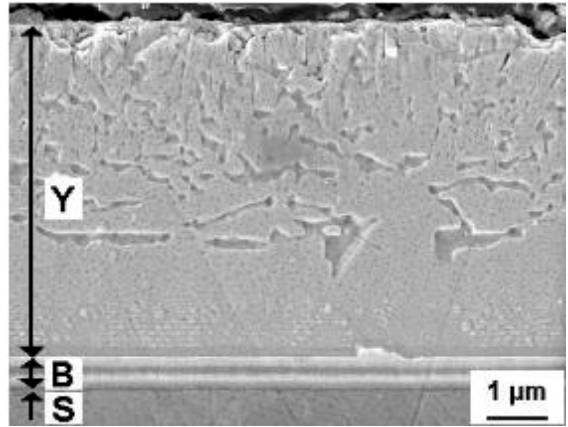

**Figure 2.** Cross-section SEM image of a PLD-grown $YBa_2Cu_3O_{7-\delta}$ film (labeled "Y") on an IBAD buffer layer ("B") on an Inconel 625 substrate ("S"). Porosity is clearly observed except for the initial ≈ 2 μm of the material. Micrograph courtesy of T. Holesinger, Los Alamos National Laboratory.

S. Foltyn and co-workers (Los Alamos National Laboratory)[70] have produced a 1 m long tape carrying $I_c = 189$ A (1 cm tape width) at 75K, corresponding to $J_c \approx 1$ MA/cm$^2$. Researchers at the University of Göttingen (H. Freyhardt *et al.*) report $I_c = 142$ A (0.92 cm tape width) on a 1.9 m long tape, corresponding to $J_c = 1.23$ MA/cm$^2$.[71]

To date, PLD appears to be the most mature technique for the growth of $YBa_2Cu_3O_{7-\delta}$ films with thicknesses above 2 μm. Results from Los Alamos National Laboratory have indicated that porous growth occurs beyond a critical thickness of about 2 μm, see Fig. 2. This agrees well with the observation of decreased $J_c$ for thicker layers. In fact, ion mill removal of the topmost portion of 3 – 6 μm thick films results in an appreciable decrease of the total critical current only when the total remaining thickness becomes less than 1.5μm (Foltyn1999b).[72]

To circumvent the problem of deteriorating microstructure in films above 2 μm thickness, Foltyn and co-workers have introduced an approach in which 0.2 μm thick layers of $SmBa_2Cu_3O_{7-\delta}$ are intercalated between separate 1 μm thick $YBa_2Cu_3O_{7-\delta}$ layers. The resulting films show a much better surface morphology (Fig.3) than standard single layers of comparable thickness, and from the achieved current densities ($J_c$ = 1.1 MA/cm$^2$) at a film thickness of 3.7 μm, a critical current in excess of 400 A/cm-width can be extrapolated. The effect of the intercalated $SmBa_2Cu_3O_{7-\delta}$ layers is most clearly visible when $J_c$ data are plotted as a function of total superconductor thickness, as in Fig. 4: For the single $YBa_2Cu_3O_{7-\delta}$ layer, the critical current does not increase beyond about 2 μm, whereas for the multilayer structures only a weak decrease of $J_c$ with thickness is observed.

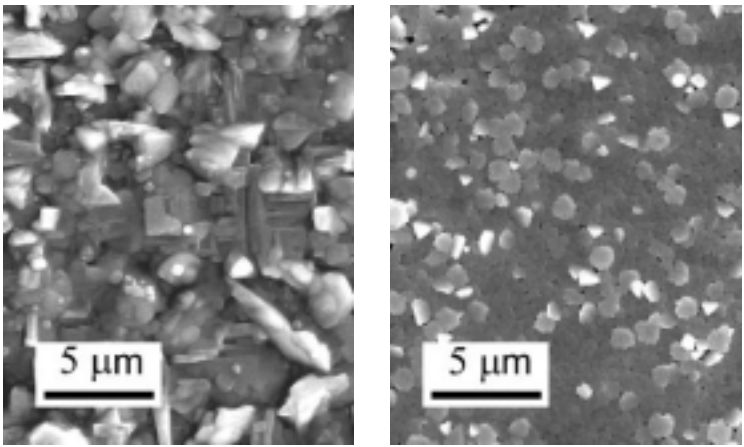

**Figure 3.** Plan-view SEM images of a 3.0 μm thick $YBa_2Cu_3O_{7-\delta}$ film (**left**) and a 3.7 μm thick $YBa_2Cu_3O_{7-\delta}$ / $SmBa_2Cu_3O_{7-\delta}$ multilayer (**right**), showing the improved surface morphology of the trilayers. Micrographs courtesy J.F. Smith, Los Alamos National Laboratory.

---

[*] Unless otherwise noted, all data quoted in this section are obtained on $YBa_2Cu_3O_{7-\delta}$ on IBAD buffer layers at 77K and in self-field.

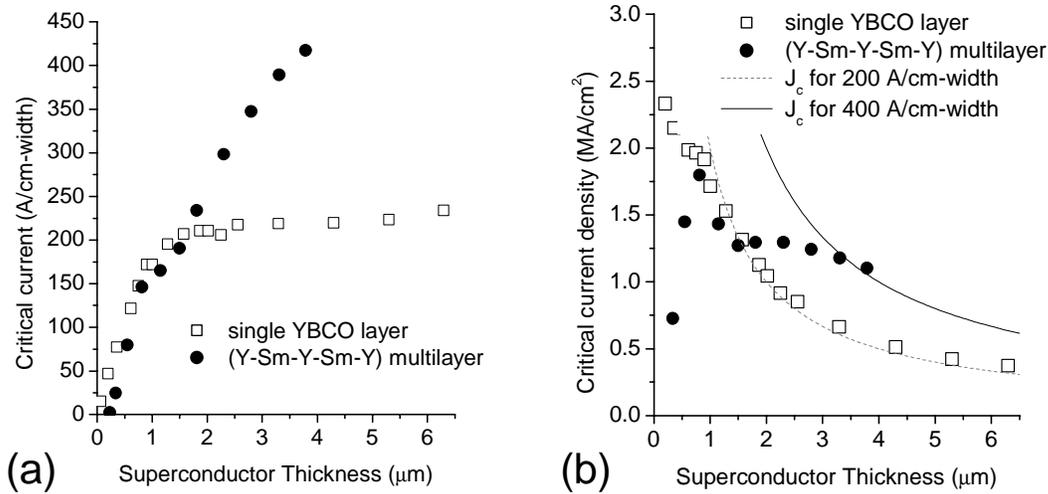

**Figure 4.** (a) Critical current (at 75K, extrapolated from measurements on 200 μm wide bridges) as a function of superconductor thickness, both for single $YBa_2Cu_3O_{7-\delta}$ films and multilayers containing two $SmBa_2Cu_3O_{7-\delta}$ films (data from Ref. 70). (b) Data from (a) plotted in terms of critical current density as function of superconductor thickness. Also shown are lines corresponding to the critical current densities required to achieve critical currents of 200 A/cm-width and 400 A/cm-width.

## COMMERCIALLY AVALIABLE EQUIPMENT

There are numerous companies worldwide selling PLD equipment, including vendors that have been focusing on PLD for many years[73] (Neocera, Inc., for example, has sold over 55 systems since 1992). Other companies offer PLD systems and components as part of a broader product line.[74] Most of these systems are research-scale tools, allowing for the coating of wafers up to a few inches in diameter. It is worthwhile to note that numerous patents have been issued on the process of pulsed laser deposition[75-80] and on various improvements thereof, including approaches for scale-up, beam scanning, in-situ monitoring, etc., some of which are referenced elsewhere in this chapter. To the best of the author's knowledge, reel-to-reel PLD systems have not been offered commercially; however, at least two companies are currently working on the development of such tools, namely PVD Products (design includes an in-situ sputter source and in-situ annealing capability) and Neocera, Inc. (developing systems for both pulsed laser and pulsed electron deposition).

Excimer laser applications such as micromachining, UV lithography, thin film transistor annealing, laser marking, fabrication of fiber Bragg gratings, etc., have created a continuous demand for reliable, industrial lasers. In addition, medical applications (including, for example, refractive laser surgery) also rely on high-quality lasers.

Modern excimer lasers, such as Lambda Physik's STEEL 1000 (Fig. 5) can produce 300 W of optical power (300 Hz repetition rate and 1 J / pulse). Designed for firing 20 million pulses a day – more than the usage of many lasers in research laboratories over the time of several months – these instruments are capable of providing the optical power required to grow $YBa_2Cu_3O_{7-\delta}$ on long lengths of tape.

## ISSUES RELATED TO SCALE-UP

### Large-area deposition

Numerous approaches to scale the process of PLD from the original sample size of a few $cm^2$ to that of wafers with diameters of several inches have been described in the literature. Careful positioning of the laser plume with respect to the center of a rotating substrate, or scanning of the laser beam across a large target[81] have been successful in uniformly coating wafers of 3" diameter reproducibly.[51] As an alternative to the scanning of a beam across a target, two laser beams can be used to form two plumes such as to create a region of sufficiently uniform particle flux.[82] Finally, the use of cylindrical targets, a long line-shaped laser spot, and a plume direction that is not normal to the substrate, combined with a substrate translation inside a radiative "pocket" heater, has been successful in depositing uniform coatings over areas as large as 7 cm x 28 cm.[38]

For the deposition of thin film coatings on tapes, however, the conditions are quite different from the situation encountered in the fabrication of films on large wafer substrates. The naturally occurring ellipsis-shaped deposition rate profile (typically measuring several centimeters along the long axis, but only 1 – 2 cm in the short direction depending on the target-substrate distance) is in fact quite appropriate for the coating of narrow tapes, as illustrated in Fig. 6.

Alternatively, the tape can be wound around a support rod in a helical shape, and this assembly can be heated by placing it inside a cylindrical black-body heater having only a relatively small opening through which deposition occurs (see Fig. 7).[71] For sufficiently rapid motion of the tape support assembly, the temperature will drop only insignificantly at the exposed deposition area. A further advantage of this approach (labeled "High-Rate Pulsed-Laser-Deposition" by the authors[83]) is the large effective deposition area: the integral deposition rate (deposited volume per unit time) can be large while the local growth rate (thickness per unit time) is kept comparatively small. This may result in increased film quality under certain conditions, but the approach may not be applicable to the growth of tapes beyond a few meters in length. A reel-to-reel system utilizing this "quasi-equilibrium heating" is currently under development.[84]

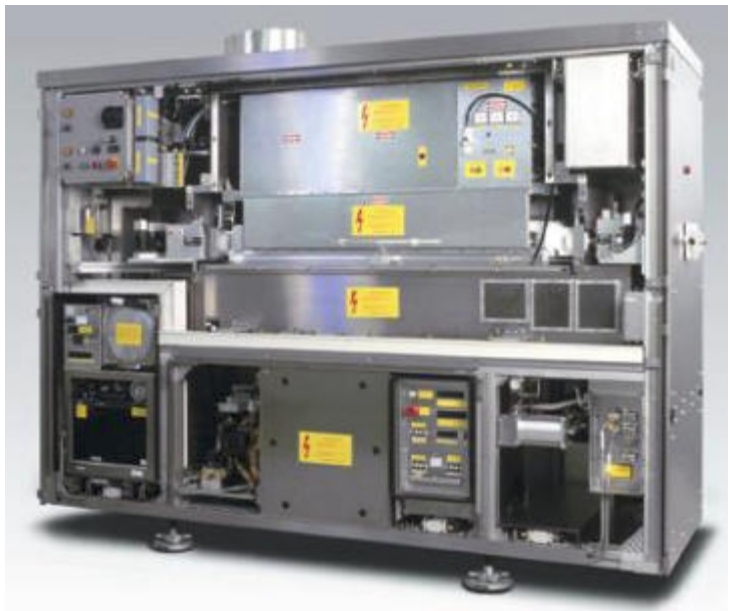

**Figure 5.** Photons at a bargain price and in a compact package: Lambda Physik's LAMBDA STEEL™ 1000 excimer laser delivers a peak optical power equivalent to that of 10 billion laser pointers. It measures 2.5 m x 2.2 m x 0.85 m (l x h x w). Photo courtesy of Lambda Physik.

**Control of the laser spot on the target**

Because the laser energy density at the target strongly influences the ablation mechanisms, the plume propagation, and the generation of particulates, a laser spot with a uniform spatial energy distribution is usually desired. In the ideal case of a perfectly parallel laser beam, a single focusing lens would be adequate to achieve this goal reasonably, considering that a sufficiently uniform fraction of the laser beam can be selected using an aperture. In reality, however, the laser beam is divergent, and the divergences in the horizontal and vertical directions are different from each other.

Using a long beam path and imaging an aperture placed nearest to the laser can result in very uniform and reproducible spots on the target. Unfortunately, the length of the required beam path for such a configuration can be quite long. For example, assume that an energy density at the target of 2 J/cm$^2$ is needed and that an energy at the laser of 800 mJ over an area of 1 cm x 3 cm (portion of the beam passing an aperture nearest to the laser) is available. Considering losses of 8% at the imaging lens and the laser port to the chamber, the remaining 677 mJ need to be focused onto a spot of about 1 mm x 3 mm, i.e. requiring a linear reduction of a factor of 10. If the distance between the target and the laser port is 50 cm, a $f = 50$ cm focal length lens can be used (resulting in a distance of the lens to the target $l_2 = (1 + 1/10) f = 55$ cm). The distance between the lens and the aperture then has to be $l_1 = (1 + 10) f = 5.5$ m.

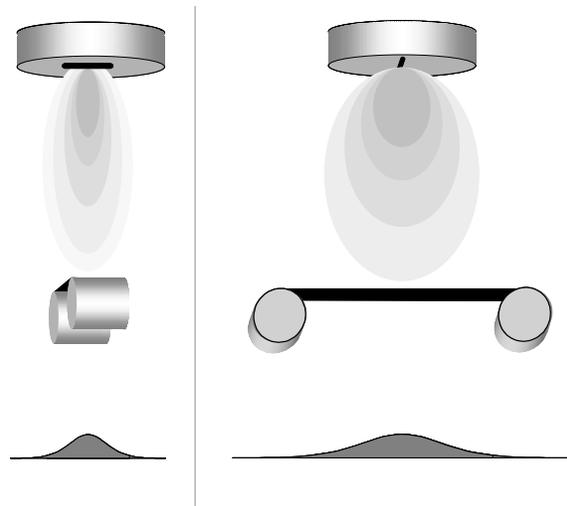

**Figure 6.** View from two orthogonal directions of a PLD plume resulting from a line-shaped laser spot, with a schematic representation of the growth rate profile as a function of position on the tape. The plume expansion is largest in the direction in which the laser spot dimension is smallest.

To gain even better spatial energy uniformity, a beam homogenizer can be used. Typically, beam homogenizers utilize an array of small (millimeter-sized) lenses to split the beam into an array of "beamlets," which are then superimposed at the focal plane directly at the target or imaged again using an additional lens (and requiring an additional length of beam path). Uniform growth has been achieved by using this approach,[38] and improved target wear has been observed.[85]

**Particulate reduction and target wear**

A number of approaches have been proposed[86,87] to reduce particulates in PLD films for devices and optical applications where perfectly smooth surfaces are often required.

The simplest approaches are arrangements in which there is no line-of-sight between the substrate surface and the laser spot on the target, resulting in deposition only via gas phase collisions while "ballistic" particulates can't reach the substrate surface. Off-axis PLD[37,88,89] is probably the simplest of these approaches. Alternatively, an "eclipse method"[35] and variations thereof[90] have been proposed, in which a small shield is placed between the target and the substrate, and film growth can again occur only due to vapor transport.

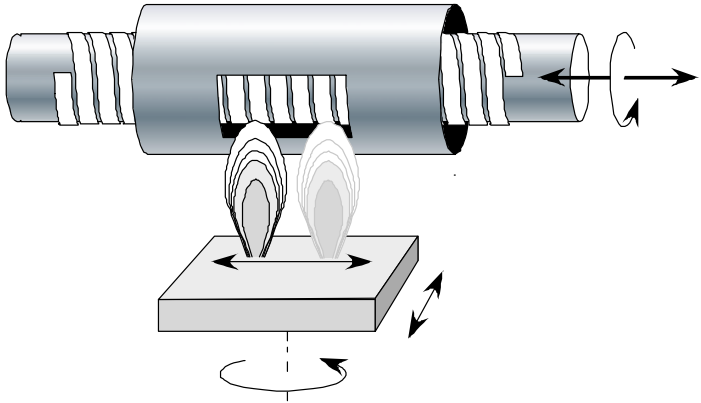

**Figure 7.** Schematic of the deposition apparatus used in Refs. 71 and 83. Target rastering and rotation is combined with beam scanning. The tape is wrapped in a helical fashion around a support rod which in turn is translated and rotated inside a heated cavity. Deposition occurs through a small opening in this heater.

In a method called "crossed-beam pulsed laser deposition,"[91,92] two laser beams are fired onto two targets that are placed at angles in such a way that the two plumes cross each other about halfway between the targets and the substrate. Beyond this crossing point, gas phase collisions lead to a plume propagation predominantly in the direction parallel to the symmetry plane of the configuration, whereas the majority of the particles does not get deflected. This approach should not be confused with a different type of "crossed beam" configuration where a pulsed gas jet is used to promote reaction in the plume,[93,94] and which also leads to a reduction in particles for appropriate timing between the valve and the laser.[86]

A different type of dual-laser approach can be used to reduce particulate *formation* (rather than to eliminate particulates once they're formed).[15] Here, a $CO_2$ laser (500 ns pulse duration) is used to first locally melt the target, which is then ablated using a KrF excimer laser pulse. Note that this method, which was first demonstrated for simple oxides, does not result in decreased deposition rates and has more recently been applied to compound semiconductors,[95] but results for $YBa_2Cu_3O_{7-\delta}$ are not available. Alternatively, a second laser can also be used to vaporize particulates in the plume.[96]

Unfortunately, all of these approaches lead to a reduced total deposition rate or an increased requirement of laser power (thus to an decreased deposition rate per unit of laser power).

Closely related to the issue of particulates is that of <u>target wear</u>. Non-uniform target erosion not only leads to changes in deposition rates and inefficient use of the target material. Most importantly, it has long been recognized that the formation of cones at the target surface[97] is a significant contribution to the particulates observed on PLD-grown films. The formation of such cones, which eventually break off and get transported onto the substrate, is most pronounced if the laser incidence angle remains constant for a large number of pulses. This can be avoided simply by a combination of target rotation with either beam scanning or target translation.

**Window coating**

A significant fraction of the laser energy can be absorbed by material accidentally deposited onto the laser entry port of the PLD system, resulting in changes in process conditions or – when properly accounted for – in increased laser energy consumption. Therefore, for large-scale depositions, care must be taken to minimize this accidental coating. Combining careful positioning of baffles within the chamber with approaches such as "window purging" (i.e. the introduction of the process gas near the laser port[98]) or additional differential pumping of the area nearest to the window[99] may be sufficient. Alternatively, active devices such as PVD Products' "Intelligent Window" have been introduced, which periodically replace the coated area of the window with a clean portion.

**In-situ monitoring and diagnostics**

Determining the composition, energetics, and dynamics of the plasma plume using a variety of techniques, including optical emission and absorption spectroscopy, time-resolved imaging, and ion probes, has been the subject of a number of studies (see Ref. 11 for a review).

Of particular importance to the fabrication of coated conductors are methods that allow real-time monitoring of certain process parameters, in particular to assure a uniform film thickness over long lengths. It is clear from the above remarks regarding window coating that a feedback mechanism is required that controls either the laser parameters (energy, repetition rate) or the tape travel speed during a long deposition.

One possible approach is to rely on an automated periodic measurement of the laser energy inside the deposition chamber to control both the laser energy and the operation of an "Intelligent Window" (as manufactured by PVD Products).

In order to determine changes both in laser energy and in the target surface condition, measurements on the growing film or on the laser plume are required. The thickness of the deposited film can be determined using ellipsometric techniques,[100,101] Raman measurements,[102] or optical spectroscopy,[103] and the amount of material deposited onto a reference point can be determined using a quartz micro-balance system.[104] Possible techniques to monitor the laser plume directly include magnetic probes,[105] ion probes,[11] and measurements of the optical emission from the plume[106,107] combined with a fully automated window actuation and laser energy control program.[108]

**SIMPLIFIED COST MODEL**

**Introduction**

It is beyond the scope of this chapter to develop a complete cost model for the PLD fabrication of coated conductors. However, PLD has gained a reputation as a somewhat costly method due to the high price of excimer lasers and the high price of the gases required. A simple calculation, however, shows quickly that neither the capital cost of an excimer laser nor the excimer gases are the dominant contribution to the cost of depositing YBCO – the cost of replacing laser tubes and optics, however, plays a significant role.

Of interest here are considerations that compare the cost of YBCO deposition with PLD to other techniques. We can therefore ignore all aspects related to tape handling, buffer layer deposition, cap layer deposition, etc., and any other aspect that is not specific to the laser operation. We further ignore the cost of ceramic targets, which are also needed in sputtering and pulsed electron deposition approaches. These ceramics may be more expensive than the materials used in evaporation; however, the forward-oriented nature of PLD results in a more efficient use of the material.

It is important to note that the numbers presented here are estimates. The high estimate will be the cost under currently established procedures, and the lowest estimate will be the best-case scenario beyond which incremental improvements of equipment and processes will not be sufficient. One must keep in mind that not all of our assumptions may be applicable to the fabrication of thousands of kilometers of tape, but also that there may be future quantum-leap improvements that we cannot predict at this time.

**Laser cost per kWhr of optical output**

We start our calculation with a determination of the cost to produce a kWhr of optical laser power, assuming the use of a Lambda Physik LAMBDA STEEL™ 1000 excimer laser. We further assume that we operate the laser 20 hours per day, 7 days per week, and 50 weeks per year. This corresponds to 21.6 million pulses per day, or $7.56 \cdot 10^9$ pulses per year at 1 J, i.e. 2100 kWhr of optical energy per year.

Lambda Physik has kindly provided the author with a cost calculation spreadsheet, on which we will base the following considerations. All calculations are made for the KrF line at 248 nm (the XeCl line at 308 nm results in operation costs that are about 2% higher due to the higher price of the gases).

A complete cost model (cost of ownership and cost of operation) takes into consideration both fixed costs (capital, floor space, static life-time of components) and variable costs (consumables, dynamic life-time of components) and considers yield and equipment duty cycle. (Definitions of cost of ownership are given and established by SEMATECH[109] and have been applied previously to excimer lasers in lithography.[110])

Current industrial lasers are very reliable, and in an early phase of tape production they are very unlikely to be a significant contribution to the overall downtime. Similarly, while there may be numerous parameters that are likely to contribute to yield issues (target changes, heating variations), the laser can be considered to be one of the more constant factors. Thus, for the costs related to the laser, yield and downtime issues may safely be ignored.

**Fixed costs.** For the purpose of the current cost comparison, we have to ignore those costs that are similar for all techniques of coated conductor fabrication. This holds in particular for facilities' cost and floor space (the foot-print of the LAMBDA STEEL™ 1000 is 2.5m x .85m, requiring a floor space of 5 $m^2$ for access and maintenance).

Other fixed costs such as static lifetime of laser gases are negligible when the laser is operated at a high duty-cycle.

Therefore, the only fixed cost that enters into our calculation is the cost of capital. According to information provided by Lambda Physik, the expected laser life is likely to be longer than the technology cycle. In other words, while it is likely that a currently available excimer laser, such as the LAMBDA STEEL™, will be replaced by newer models in a few years, the actual life expectancy of the laser is much higher (10 to 20 years or more) if properly maintained.[111] Note that in our considerations for variable costs, we will consider items such as laser tubes and optics, which have a life expectancy that is measured in laser pulses. However, other components such as electrical power supplies, fan motors, cooling components, etc., are not considered.

Assuming, therefore, a laser life of 5 to 20 years, and an estimated laser price of $800k, the capital cost per laser is between about $50k (optimistically assuming 3% annual interest and a 20 year depreciation) and $200k (pessimistically assuming 8% interest and a 5 year depreciation). Thus, the capital cost per kWhr is about $20 - $100.

**Variable costs.** According to the information provided by Lambda Physik, at the above-mentioned duty cycle, 484 gas fills are required per year, the window sets need to be exchanged 75 times, the thyrotron modules 10 times, and the laser tube between 3 and 4 times annually. Estimated costs for the window sets are about $128k per year and for the laser tubes $280k per year. In comparison, the laser gases cost only about $19k per year, 75% of which is for Neon. In total, Lambda estimates a cost of $480k per year for replacement parts and consumables, i.e. $228/ kWhr.

Other variable costs are utilities and labor. Lambda estimates 212 hours of required labor per year. At a fully burdened technician cost of $80k to $160k, this adds about $4 – 8/kWhr of laser output.

The cost of the required electricity is harder to estimate. First we note that the electrical power consumption of the laser is on the order of 30 kVA, and that 20 kW of cooling capacity (using cooling water at 10°C) are required. Furthermore, an air flow of 1000 m$^3$/hr is needed. While it is beyond the scope of this chapter to estimate the efficiency of the water cooling or the price of electricity 5 years from now, assuming 40 - 60 kW of power consumption at $0.03 to $.1/kWhr (Ref. 112) adds $8k - $42k annually or $4 - 20 per kWhr of laser output.

**Total cost of laser energy.** Figure 8 summarizes the various fixed and variable cost components as describe above. It can therefore be estimated that the total cost of the optical energy falls into the range of $250/kWhr – $350/kWhr. It is impossible to predict if these costs will decrease significantly in the foreseeable future. Past cost savings have, for example, resulted from the introduction of Lambda Physik's NovaTube™ technology. A similar large change in laser technology would, however, be required for future appreciable price reductions.[111]

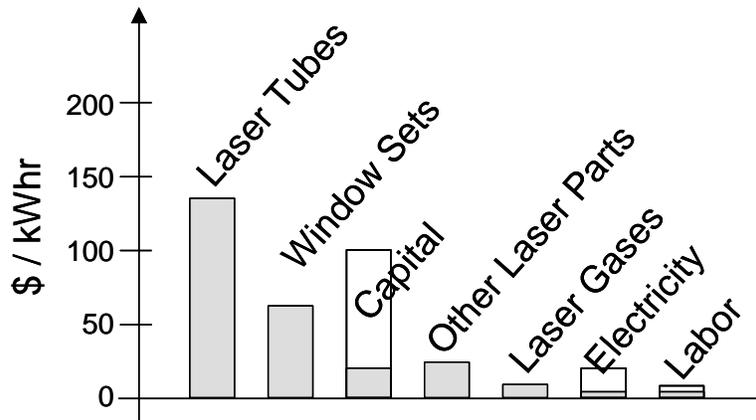

**Figure 8.** Fixed and variable costs considered for a LAMBDA STEEL™ 1000, resulting in a cost of $250 - $350 per kWhr of optical output.

**Optical energy required to grow a meter of tape**

It is more difficult to estimate how much optical energy is required to grow a certain length of superconducting tape. However, there are a number of published reports that indicate the current status. Foltyn *et al.* reported the growth of 1 μm thick $YBa_2Cu_3O_{7-\delta}$ at a rate of 2.5 cm/min. using 14 W of excimer laser power,[113] corresponding to 0.0093 kWhr/m. Using a 200W laser, the same authors later reported a growth rate of 17 m/hr, corresponding to 0.011 kWhr/m,[114] indicating in fact that the process scales almost perfectly with higher laser power and repetition rate. Iijima *et al.* predict that the growth of 3 – 4 m per hour with a thickness in excess of 1μm should be possible with less than 100 W of optical power,[115] corresponding to less than 0.025 kWhr/m.

In a very different configuration, the researchers at the university of Göttingen report a growth rate of 45 nm m$^2$/hr, using a LAMBDA 3308 excimer laser at 300 Hz and 0.4 J/pulse. These numbers correspond to 0.027 kWhr/m for a 1cm wide, 1 μm thick tape.

Several factors have to be considered when estimating the minimum laser power required for film growth. First, current implementations of PLD are typically not optimized for the lowest loss in laser energy. For the values from Los Alamos cited above, for example, it is estimated[116] that at least 25% in the laser energy on the target could be gained by using optimized optical elements and a shorter beam path.

A further significant factor is the shape of the laser spot and thus the shape of the plume, and the amount of material that gets deposited onto the tape rather than outside of the tape. While PLD is extremely forward-oriented when compared to some other methods, it is obvious that for cost-efficient manufacturing, some of the currently achieved film thickness uniformity can be traded in for an increased integrated collection efficiency. According to S. Foltyn,[116] one can expect to gain up to a factor of two in the deposition rate at given laser power by optimizing beam path, optical elements, laser spot, and growth geometry.

Therefore, we can assume that the energy required to deposit one meter of tape (with a thickness of 1 µm and a width of 1 cm) lies between 0.005 kWhr (assuming a two-fold increase in overall efficiency due to increased beam path and collection efficiencies) and 0.008 kWhr (assuming only a 25% increase in the beam path efficiency).

**Total estimated laser cost per length of tape**

To summarize the numbers calculated above, we find that
- 1 kWhr of optical energy costs about $250 – $350.
- The growth of 1 m of superconducting tape with a width of 1 cm and film thickness of 1 µm requires about 0.005 kWhr – 0.008 kWhr (with currently achieved numbers sometimes significantly higher due to non-optimized beam paths and deposition geometries).

Using these numbers, the laser cost per meter of tape (1 cm wide, 1 µm thick) can be calculated to fall into the range of $1.2 - $2.8. This corresponds to a laser operation cost of $12 - $28 per kA m if a critical current density of 1 MA/cm$^2$ is assumed. Costs for other values of $J_c$ and energy consumption can be determined from the plots in Fig. 9.

To summarize, even with existing technology, the laser cost is less than $30/kA m, but any currently discussed incremental improvements of equipment and process is very unlikely to result in a laser cost of less than $10/kA m for $J_c$ = 1 MA/cm$^2$ (this cost, obviously, varies as $J_c^{-1}$).

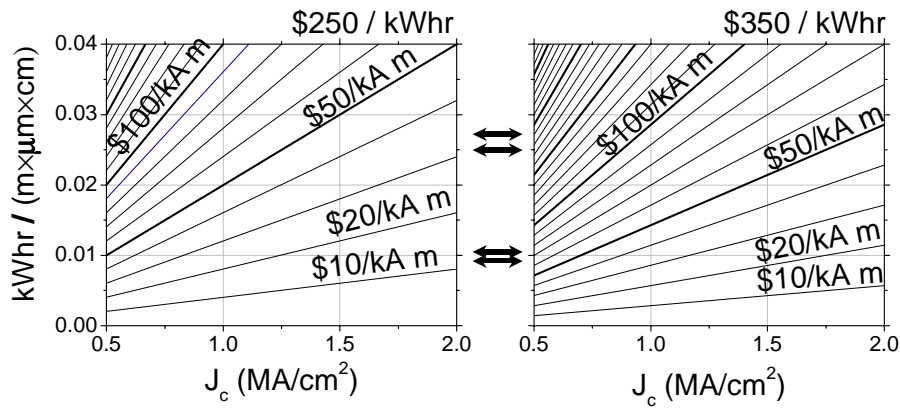

**Figure 9.** Laser cost contour plots for an assumed cost of $250 (left) and $350 (right) per kWhr of optical energy. Both the expected $J_c$ (x-axis) and the expected laser energy consumption per length of tape (y-axis) must be considered to determine the actual cost per kA m. The double arrows between the plots indicate the experimental values quoted in the text.

## SUMMARY AND CONCLUSIONS

Pulsed laser deposition has been shown to be a powerful method for the fabrication of coated conductor tapes. Some of the highest values of $I_c$ reported have been obtained by PLD, as this technique has been most successful in the growth of $YBa_2Cu_3O_{7-\delta}$ films with thicknesses exceeding 1 – 2 μm.

What are often described as the main disadvantages of PLD, namely the non-uniform deposition profile and the formation of particulates, appear to be much less of a concern in the fabrication of coated conductors than in thin films for device applications. In addition, many of the critical aspects for scale-up, such as laser port coating, target wear, and in-situ monitoring of deposition rates, have already been addressed successfully.

Unfortunately, excimer lasers are relatively expensive to operate. Using the current state of PLD, the cost of the required photons alone may contribute up to \$30/kA m. Incremental improvement of the process (optimized laser beam paths and collection efficiencies) may result in a decrease of the cost to just above \$10/kA m, assuming a $J_c$ of 1 $MA/cm^2$. A significant further decrease of the cost would require quantum-leap changes in laser technology, deposition method, or achievable $J_c$.

## ACKNOWLEDGEMENTS

It is my great pleasure to acknowledge numerous fruitful conversations with S.R. Foltyn and V. Matijasevic (Los Alamos National Laboratory), Y. Iijima (Fujikura Ltd.), H.C. Freyhardt (University of Göttingen), K.S. Harshavardhan and G. Doman (Neocera, Inc.), J.A. Greer (PVD Products), J. Maclin (Lambda Physik), D.H.A. Blank (University of Twente), P.K. Schenck (NIST Gaithersburg), and D.B. Geohegan, G. Eres, C.M. Rouleau, A. Puretzky, and D.H. Lowndes (Oak Ridge National Laboratory). This work was sponsored by the U.S. Department of Energy under contract DE-AC05-00OR22725 with the Oak Ridge National Laboratory, managed by UT-Battelle, LLC, and by the DOE Office of Energy Efficiency and Renewable Energy, Office of Power Technologies – Superconductivity Program.